\documentclass[twocolumn,showpacs,preprintnumbers,amsmath,amssymb,aps]{revtex4}

\usepackage{bm}
\usepackage{graphicx}
\usepackage{bm}

\begin{document}

\title{Hamiltonian of a many-electron system with single-electron and electron-pair states 
in a two-dimensional periodic potential}

\author{G.-Q. Hai$^1$ and F. M. Peeters$^2$}

\affiliation{$^{1}$Instituto de F\'{\i}sica de S\~{a}o Carlos, Universidade de S\~{a}o Paulo, 13560-970, S\~{a}o Carlos, SP, Brazil}
\affiliation{$^{2}$Departement Fysica, Universiteit Antwerpen, Groenenborgerlaan 171, 
2020 Antwerpen, Belgium}

\begin{abstract}
Based on the metastable electron-pair energy band in a two-dimensional (2D) periodic potential 
obtained previously by Hai and Castelano [J. Phys.: Condens. Matter 26, 115502 (2014)],
we present in this work a Hamiltonian of many electrons consisting of single electrons 
and electron pairs in the 2D system. The electron-pair states are metastable of energies 
higher than those of the single-electron states at low electron density. 
We assume two different scenarios for the single-electron band. When it is considered 
as the lowest conduction band of a crystal, we compare 
the obtained Hamiltonian with the phenomenological model Hamiltonian of a boson-fermion 
mixture proposed by Friedberg and Lee [Phys. Rev. B 40, 6745 (1989)].
Single-electron-electron-pair and electron-pair-electron-pair interaction terms appear in 
our Hamiltonian and the interaction potentials can be determined from the electron-electron Coulomb interactions. When we consider the single-electron band as the highest valence band of a crystal, 
we show that holes in this valence band are important for stabilization of the electron-pair states in the system. 
\end{abstract}



\maketitle

The mechanism of electron pairing in unconventional superconducting materials still remains 
as one of the major and unresolved problems in condensed matter physics.
It is usually believed that the 2D characteristic (such as a layered crystal structure of 
the cuprate superconductors) and strong electron-electron correlation are important for unconventional superconductivity.\cite{sci,aks,prl2011}
From the experimental observation of the very small coherence length in hight $T_c$ superconductors
Friedberg and Lee\cite{FL} (referred to as FL below) proposed a phenomenological theory, the so-called $s$-channel theory. 
In this theory of the boson-fermion model, they introduced phenomenologically a local boson 
field $\phi$ representing a pair state which
is localized in coordinate space. Each individual $\phi$ quantum
is assumed to be unstable with an excitation energy $2\nu$. It has been shown that this 
assumption makes it possible for the $s$-channel theory to exhibit many superconducting characteristics\cite{FL,FLR,Ren,GIL97,MDell,EP2003}. 
This model has been studied both for three-dimensional (3D) and two-dimensional (2D) systems. 
As a matter of fact, the boson-fermion model has been extensively studied in the area of 
unconventional superconductivity.\cite{FL,FLR,Ren,GIL97,MDell,EP2003,MRS1954,RMR,RRE,prl2013} 
It can be traced back before the BCS theory for conventional superconductivity.\cite{MRS1954}

The mechanism of electron pairing and the nature of the bosonic excitations
are still key issues for the understanding of unconventional superconductivity.\cite{Sci355}
Based on previously obtained metastable electron-pair states in a 2D crystal\cite{HC},
we present in this work a Hamiltonian of a 2D electron system consisting of single electrons (fermions) and electron pairs (bosons). This new Hamiltonian obtained from electronic structure
calculations of a 2D crystal includes single-electron states, electron-pair states, as well as Coulomb interactions between single electrons and electron pairs. 
The interaction potentials present in the Hamiltonian can be obtained from 
the electronic structure calculations with electron Coulomb interaction. 

We start with the single-electron states in a 2D square lattice crystal. The crystal periodic
potential is given by  $V_c(x,y)=V_0[\cos(q x)+\cos(q y)]$, where $q=2 \pi / \lambda$ with $\lambda$ 
the period and $V_0$ the amplitude of the crystal potential. 
The Schr{\"o}dinger equation for a single electron is given by 
$H({\bf r}) \psi_{n,{\bf k}}({\bf r})=E_{n,{\bf k}} \psi_{n,{\bf k}}({\bf r})$ with 
the single-electron Hamiltonian $H({\bf r}) = - \nabla ^2 +V_c(x,y)$ in the units of effective 
Bohr radius ${\rm a}_B$ and effective Rydberg ${\rm R}_y$, 
where ${\bf k}$ is the wavevector in the first Brillouin zone, $n$ is the band index, 
$E_{{n,\bf k}}$ and $\psi_{n,{\bf k}}({\bf r})$ are the eigenvalue 
and eigenfunction, respectively. 

When we consider two interacting electrons in this periodic potential, the Hamiltonian is given by 
$H({\bf r}_1,{\bf r}_2 ) =H ({\bf r}_1)+ H ({\bf r}_2) +2/|{\bf r}_1 - {\bf r}_2 |$.
The ground state of this two-electron system can be found for $|{\bf r}_1 - {\bf r}_2| \to \infty $ due to the electron-electron repulsion. It means that the ground state consists of two non-interacting single electrons separated by an infinitely long distance.
However, it was found previously that metastable electron-pair states\cite{HC}
may exist in this system. In such an electron-pair state, two electrons of spin singlet stay in 
the same unit cell of the crystal lattice in the relative coordinates ${\bf r}={\bf r}{_1}-{\bf r}_2$ with an average distance between them only about one third of the lattice period. Because two electrons  in the same unit cell are of maximum overlap, they are strongly correlated. Strong electron-electron correlation reduces greatly the Coulomb repulsion between them. When the amplitude 
$V_0$ of the crystal potential is larger than a certain value, strong correlation between two electrons together with the crystal potential modulation leads to a metastable electron-pair 
state. The electron-pair states in the center of mass coordinates 
{\bf R}=(${\bf r}_1+{\bf r}_2$)/2 are 
given by Bloch wavefunctions and form energy bands. The eigenenergies of the electron pairs are 
given by $E^{\rm pair}_{\bf k}$ and the wavefunctions can be written as 
\[
\Psi_{\bf k}({\bf R},{\bf r})
=\frac{e^{i{\bf k}\cdot{\bf R}}}{\sqrt{A}}\sum_{l_x l_y;nm} a_{l_xl_y;nm} 
e^{i{\bf G}_l\cdot {\bf R}}  R_{nm}(r)\frac{\cos(m\theta)}{\sqrt{b_m\pi}},
\]
where ${\bf G}_l=l_x q{\bf i}+l_y q {\bf j}$ (with $l_x, l_y= 0, \pm1, \pm2,...$) is the reciprocal-lattice vector, $R_{nm}(r)$ (with $n$=0,1,2,..., and $m$=0,2,..., n) is taken from the eigenfunction of a hydrogen atom with a scaling factor of Laguerre function as given in Ref.\cite{HC}, $b_0=2$ and $b_m=1$ for $m \geq 1$. 
It was found that the lowest eigenvalue of an electron pair is spin singlet and forms an energy band. In terms of the energy per electron, this electron-pair band is found in between the two lowest single-electron bands. The minimum of the electron-pair band at the $\Gamma$ point in the first Brillouin zone is higher than the maximum of the lowest single-electron band (at the $M$ point) in the studied square lattice potential in Ref.\cite{HC}. Their difference is defined as an energy gap $E_g$.

In the presence of many electrons in the crystal, we expect that for a certain electron density 
many-body interactions may stabilize the electron-pair states leading to a macroscopic electron-pair density. Without loss of generality, we consider two energy bands: a lower band $E^{\rm s}_{\rm k}$ 
of single-electron states  $\psi_{\bf k}({\bf r})$ and a higher band $\nu_0 + E^{\rm pair}_{\rm k}/2$ (given by energy per electron) of electron-pair states. As was shown in Ref.\cite{HC}, 
the minima of the two bands in the square lattice are at the $\Gamma$ point in the first Brillouin zone and their difference at low-electron density is given by $\nu_0 =E_g+E_{BW}$, 
where $E_{BW}$ is the band width of the single-electron band. 

In order to obtain the many-electron Hamiltonian, we will consider two different 
scenarios for the single-electron band. 
We consider first the single-electron band as the lowest conduction band in a real crystal.
Free conduction electrons occupy this band. It is empty in an undoped insulator at low temperature.  
When the electron-pair band is taken into account, some electrons may form metastable pairs in 
this band. 
We assume that there are $N$ electrons consisting of $N_F$ single electrons and $N_B$ electron 
pairs in the system ($N=N_F+2 N_B$). At low electron density, 
all conduction electrons occupy the single-electron band, i.e., $N=N_F$ and $N_B=0$. At higher densities, electron pairs may appear and many-body interaction may stabilize these pairs. 
The Hamiltonian of the system can be written as,
\begin{eqnarray}\label{H}
H= && \sum_{i=1}^{N} H_0 ({\bf r}_i) + 
\frac{1}{2}\sum_{i=1}^{N}\sum_{j\neq i}^N \frac{2}{|{\bf r}_i - {\bf r}_j |} \nonumber\\
= && \sum_{i=1}^{N_F} H_0 ({\bf r}_i) + 
\frac{1}{2}\sum_{i=1}^{N_F} \sum_{j\neq i}^{N_F} \frac{2}{|{\bf r}_i - {\bf r}_j |} \nonumber\\
+ && \sum_{\alpha=1}^{N_B}\left[ \sum_{\nu=1}^2 H_0 ({\bf r}_{\alpha,\nu})+ 
\frac{1}{2} \sum_{\nu=1}^2\sum_{\mu\neq\nu}^2 \frac{2}{|{\bf r}_{\alpha,\nu}-{\bf r}_{\alpha,\mu}|}\right] \nonumber\\
+ &&
\frac{1}{2} \sum_{\alpha=1}^{N_B}\sum_{\beta\neq\alpha}^{N_B} \sum_{\nu=1}^2 \sum_{\mu=1}^2
\frac{2}{|{\bf r}_{\alpha,\nu}-{\bf r}_{\beta,\mu}|}  \nonumber\\
+ && \sum_{i=1}^{N_F} \sum_{\alpha=1}^{N_B} \sum_{\nu=1}^{2} \frac{2}{|{\bf r}_i-{\bf r}_{\alpha,\nu}|},
\end{eqnarray}
where ${\bf r}_i$ denotes single electron $i$, and ${\bf r}_{\alpha,\nu}$ electron pair $\alpha$  (with two electrons $\nu$=1 and 2). 
The first two terms in the above Hamiltonian are $N_F$ single electrons and the Coulomb interactions between single electrons. The next two terms are $N_B$ electron-pairs and the Coulomb interactions between the pairs.
The last term is the Coulomb interactions between single electrons and electron pairs.  
Due to possible transitions: \emph{two single electrons}  $\rightleftarrows$ \emph{an electron pair}, $N_F$ and $N_B$ are not constants.
Two single electrons can form an electron pair through the Coulomb interaction between them. 
In the opposite process, an electron pair can decouple into two single electrons.  
A macroscopic number of electron pairs in the system depends on the electronic band 
structure of the crystal (determined by the crystal potential), the total electron density and temperature, etc.. When there are no electron pairs in the system, the above Hamiltonian reduces to that of interacting single electrons with $N=N_F$. 

Using the eigenfunctions of the single-electron and electron-pair states, we can obtain the field 
operators of single electrons and electron pairs. 
The single electron (fermion) field operators can be written as,
\begin{equation}
\tilde{\psi}_{\sigma}({\bf r})=\sum_{\bf k} \psi_{\bf k}({\bf r}) c_{{\bf k},\sigma}
\end{equation}
and 
\begin{equation}
\tilde{\psi}^\dag_{\sigma}({\bf r})=\sum_{\bf k} \psi^{*}_{\bf k}({\bf r}) c^\dag_{{\bf k},\sigma},
\end{equation}
where the operators $c_{{\bf k},\sigma}^\dag$ and $c_{{\bf k},\sigma}$ are creation and annihilation operators, respectively, for a single electron of momentum $\hbar${\bf k} and spin ${\sigma}$. 
They obey the fermion anti-commutation relations
$\{ c_{{\bf k},\sigma}, c^{\dag}_{{\bf k}^{\prime},{\sigma}^{\prime} } \} = 
\delta_{{\bf k},{\bf k}^{\prime}} \delta_{{\sigma},{\sigma}^{\prime}} $,
$\{ c_{{\bf k},\sigma}, c_{{\bf k}^{\prime},{\sigma}^{\prime} } \} = 0$, and
$\{ c^\dag_{{\bf k},\sigma}, c^{\dag}_{{\bf k}^{\prime},{\sigma}^{\prime} } \} = 0$. 

The electron-pair (boson) field operators are given by
\begin{equation}
\tilde{\Psi} ({\bf R},{\bf r})= \sum_{\bf k} \Psi_{\bf k}({\bf R},{\bf r}) b_{\bf k} 
\end{equation}
and
\begin{equation}
\tilde{\Psi}^{\dag} ({\bf R},{\bf r})= \sum_{\bf k} \Psi^{*}_{\bf k}({\bf R},{\bf r}) 
b^{\dag}_{\bf k}, 
\end{equation}
where the operators $b^\dag_{\bf k}$ and $b_{\bf k}$ are creation and annihilation operators, respectively, for an electron pair of momentum $\hbar${\bf k}. They obey the commutation relations
$ [ b_{\bf k}, b^{\dag}_{{\bf k}^{\prime}} ] = \delta_{{\bf k},{\bf k}^{\prime}}  $,
$ [ b_{\bf k}, b_{{\bf k}^{\prime}} ] =0  $, and
$ [ b^\dag_{\bf k}, b^{\dag}_{{\bf k}^{\prime}} ] = 0$.

Considering that the electron-pair number $N_B= (N-N_F)/2 $ in the system depends on 
many-particle interactions and band re-normalization,  
the Hamiltonian of the system in second quantization 
with single-electron and electron-pair states can be written in two parts, 
\begin{equation}\label{Ht}
\tilde{H} = \tilde{H}_0+\tilde{H}_1 .
\end{equation}
The first part $\tilde{H}_0$  corresponds formally to the five terms in Eq.(\ref{H}), given by 
\begin{eqnarray}\label{Hpair}
\tilde{H}_0=&& 
\sum_{{\bf k},\sigma} E^s_{{\bf k}} c^{\dag}_{{\bf k},\sigma} c_{{\bf k},\sigma} \nonumber \\
+ && \frac{1}{2A}\sum_{\bf k_1,k_2, q} \sum_{\sigma, \sigma^\prime} v_{q} 
c^{\dag}_{\bf k_1 -q,\sigma} c^{\dag}_{\bf k_2 +q,\sigma^\prime} 
c_{\bf k_2,\sigma^\prime}c_{\bf k_1,\sigma}     \nonumber \\
+ && \sum_{\bf k} \left[ E^{\rm pair}_{\bf k}+2\nu_0 \right] b^{\dag}_{\bf k} b_{\bf k} \nonumber \\
+ && \frac{1}{2A}\sum_{\bf k_1,k_2, q} v_{pp}(q) 
b^{\dag}_{\bf k_1 -q} b^{\dag}_{\bf k_2 +q} b_{\bf k_2}b_{\bf k_1}  \nonumber \\
+ && \frac{1}{2A}\sum_{\bf k, k_1, q,\sigma} v^{\rm s}_{ep}(q) 
b^{\dag}_{\bf k_1 -q} c^{\dag}_{\bf k +q,\sigma} b_{\bf k_1}c_{\bf k,\sigma} ,
\end{eqnarray}
where $v_q= 2 \frac{2\pi}{q}$ is the electron-electron Coulomb potential,
$v_{pp}(q)=4 v_q f_{pp}(q)$ the pair-pair interaction potential with a form factor $f_{pp}(q)$, 
and $v^{\rm s}_{ep}(q)=  2 v_q f_{ep}(q)$ the single-electron-electron-pair interaction 
(scattering) potential with form factor $f_{ep}(q)$. Notice that, in Eq.(\ref{Hpair}) we have taken the bottom of the single-electron band as the reference for energy. The summation over {\bf q} does not include {\bf q}=0 because it cancelled out with the background ion-ion interaction and the system is neutral.     
The form factors $f_{ep}(q)$ and $f_{pp}(q)$ reflect finite size of the electron-pair
(against point charge) and are given by
\begin{equation}
f_{ep}(q) = \sum_{l_x l_y} \sum_{n m} \sum_{n^\prime m^\prime} 
a^{*}_{{l_x l_y};n^\prime m^\prime} a_{{l_x l_y};nm} 
F_{nm;{n^\prime}m^\prime}(q),
\end{equation}
and 
\begin{equation}
f_{pp}(q)= [f_{ep}(q)]^2,
\end{equation}
respectively, with
\begin{eqnarray}
 F_{nm;{n^\prime}m^\prime}(q)= && \frac{(-1)^{\frac{m+m^\prime}{2}}}{\sqrt{b_m b_{m^\prime}}} 
 \int_0^{\infty} r dr R_{nm}(r)R_{n^\prime m^\prime}(r) \nonumber\\
 && \times \left[ J_{m+m^\prime} (\frac{_{qr}}{^2}) 
 +J_{|m-m^\prime|} (\frac{_{qr}}{^2}) \right],
\end{eqnarray}
where $J_m(x)$ is the Bessel function of the first kind. 

\begin{figure}[htb!]
      {\includegraphics[width=8cm,height=5cm]{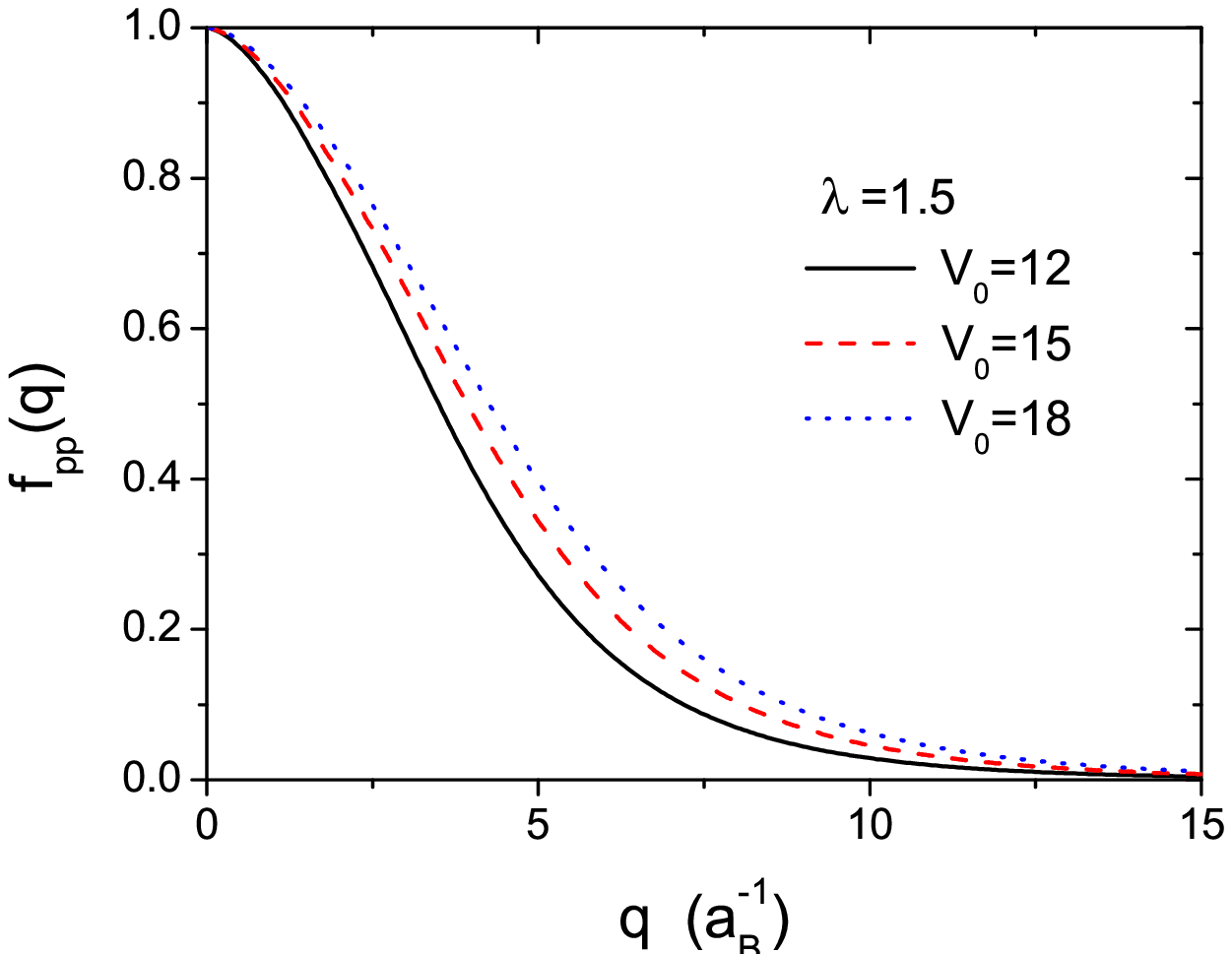}}
      {\includegraphics[width=8cm,height=5cm]{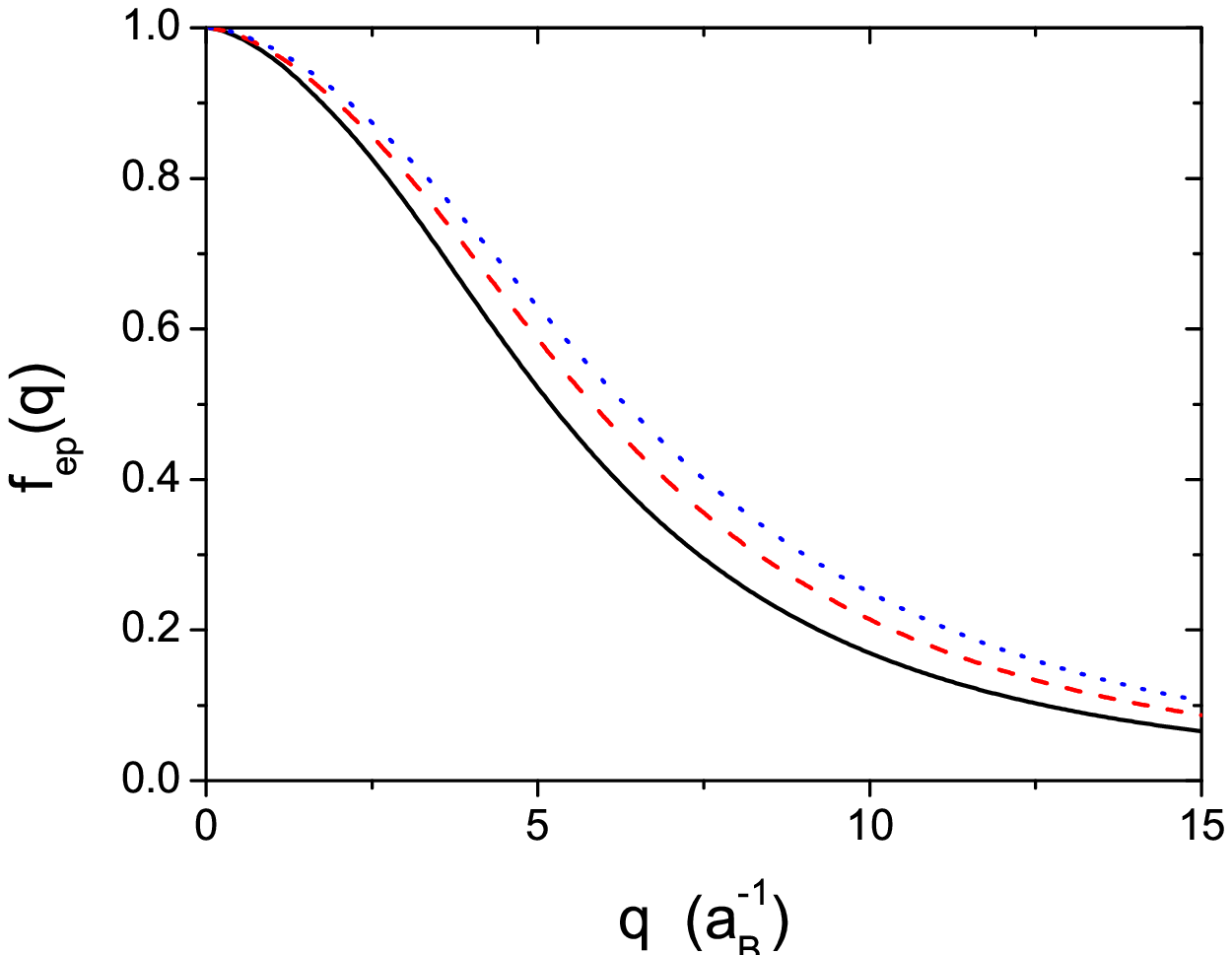}}
      {\includegraphics[width=8cm,height=5cm]{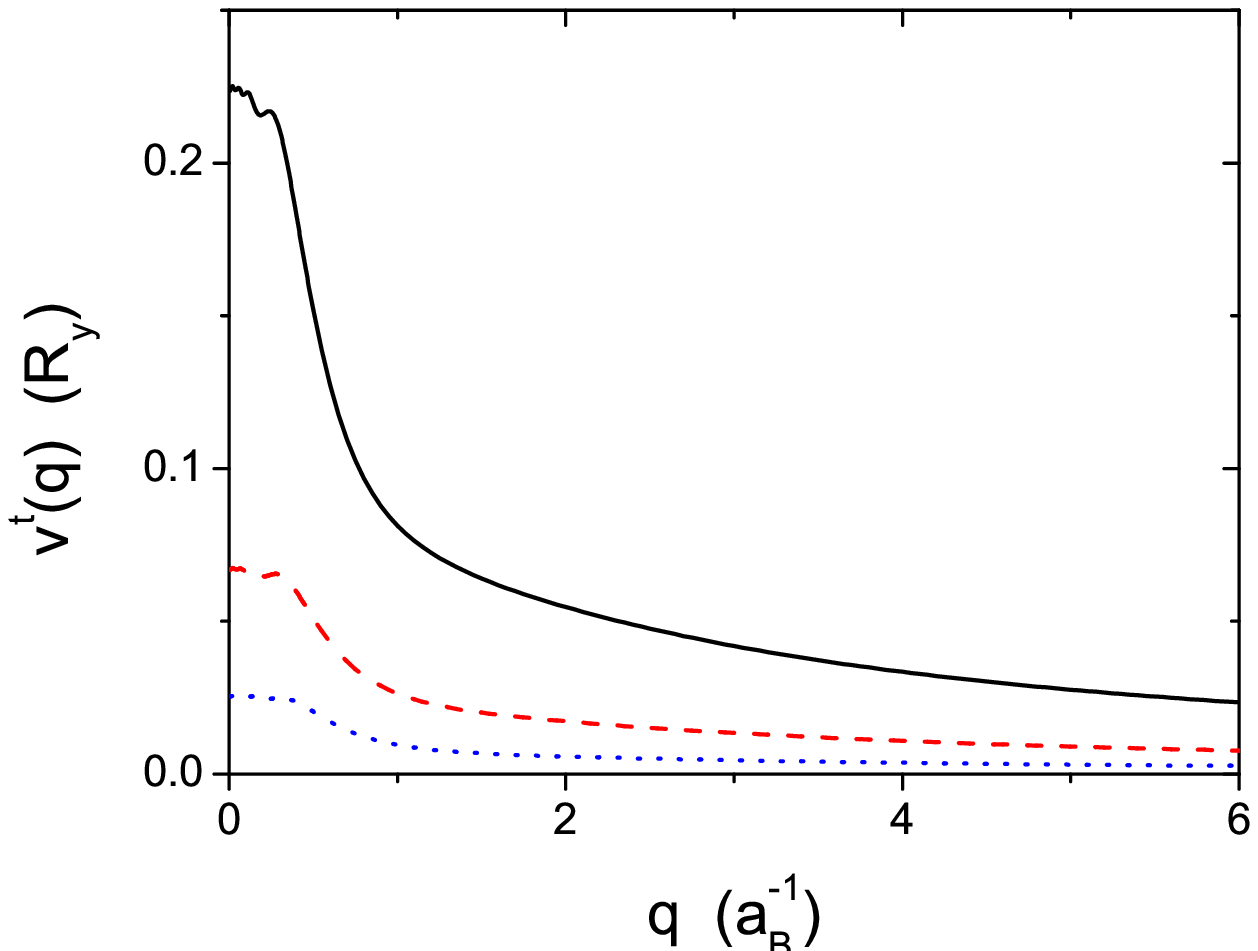}}
       \caption{The form factors for the electron-pair-electron-pair ($f_{pp}(q)$) 
       and single-electron-electron-pair ($f_{ep}(q)$) interaction potentials,
       and the interaction potential $v^{t}(q)$ for single-electrons-electron-pair transition. 
       $\lambda=1.5$ ${\rm a_B}$ and $V_0=$ 12, 15, and 18 ${\rm R_y}$.}
       \label{fig}
\end{figure}

The second part $\tilde{H}_1$ of the total Hamiltonian Eq.~(\ref{Ht}) is responsible for 
the transition or transformation between two single electrons and an electron pair. 
The interaction potential is just the Coulomb potential between the two electrons. 
It is given by  
\begin{eqnarray}\label{H1}
\tilde{H}_1&=&\int d{\bf r}_1 \int d{\bf r}_2 \tilde{\Psi}^{\dag} ({\bf R},{\bf r})
\frac{2}{|{\bf r}_1 - {\bf r}_2 |}
\tilde{\psi}_{\uparrow}({\bf r}_1) \tilde{\psi}_{\downarrow}({\bf r}_2) \nonumber\\
&+&\int d{\bf r}_1 \int d{\bf r}_2 
\tilde{\psi}^{\dag}_{\uparrow}({\bf r}_2) \tilde{\psi}^{\dag}_{\downarrow}({\bf r}_1) 
\frac{2}{|{\bf r}_1 - {\bf r}_2 |}
\tilde{\Psi}({\bf R},{\bf r})  \nonumber\\
= &\frac{1}{\sqrt{A}}&\sum_{\bf k, q} v^{\rm t}(q)
\left( b^{\dag}_{\bf k } c_{\frac{\bf k}{2}+{\bf q},\uparrow} c_{\frac{\bf k}{2}-{\bf q},\downarrow} 
+ b_{\bf k } c^{\dag}_{\frac{\bf k}{2}-{\bf q},\downarrow} 
c^{\dag}_{\frac{\bf k}{2}+{\bf q},\uparrow}  \right), \nonumber\\
\end{eqnarray}
where $v^{\rm t}(q)$ is the single-electron-electron-pair interaction potential for 
transition (transformation) which is given by
\begin{equation}\label{v-pp}
v^{\rm t}(q)=
\sum_{l_x l_y;nm} \frac{2\sqrt{\pi}(-1)^{\frac{m}{2}}}{\sqrt{b_m}} a^{*}_{{l_x l_y};nm} 
\int_0^\infty dr R_{nm}(r) J_m(qr).
\end{equation}

The form factors $f_{pp}(q)$ and $f_{ep}(q)$ and the potential $v^{\rm t}(q)$ can be obtained from numerical calculation. They are given in Fig. 1 for the square lattice with $\lambda=1.5$ a$_B$ and $V_0=12$, 15, and 18 R$_y$.

The total electron number operator is given by
\begin{eqnarray}
\tilde{N} =&&
\int  d{\bf r} \sum_{\sigma}\tilde{\psi}^\dag_{\sigma}({\bf r}) \tilde{\psi}_{\sigma}({\bf r})+
  2 \int d{\bf R} \int d{\bf r}\tilde{\Psi}^{\dag}({\bf R},{\bf r}) \tilde{\Psi} ({\bf R},{\bf r})
  \nonumber \\
  = &&\sum_{\bf k,\sigma} c^{\dag}_{{\bf k},\sigma} c_{{\bf k},\sigma}
   + 2\sum_{\bf k} b^{\dag}_{{\bf k}} b_{{\bf k}} .
\end{eqnarray}
It commutates with the total Hamiltonian Eq.~(\ref{Ht}) and is conserved.  

In comparison with the FL Hamiltonian\cite{FL}, the interaction terms due to
pair-pair and single-electron-electron-pair interactions appear in our Hamiltonian given by Eqs.~(\ref{Ht}), (\ref{Hpair}), and (\ref{H1}). 
The FL model Hamiltonian of the boson-fermion mixture was given by Eqs.(1.7), (1.8) and (1.9)
in Ref.~[\onlinecite{FL}] containing three terms only, which correspond to the single electrons  
(or holes), the single bosons (i.e., electron pairs) and the interaction between electrons (or holes) and the bosons. The boson quantum was introduced phenomenologically with spin zero and with 
twice the mass and charge of a single electron, and it was further assumed unstable with an excitation energy 2$\nu$. But, the detailed microscopic structure of the boson quantum and its origin of the pairing mechanism were unknown. Therefore, the interaction potential between the electrons and bosons was described through a coupling constant as a parameter. 
In this work, however, we build a more complete theory with the Hamiltonian given in Eqs.~(\ref{Ht}), (\ref{Hpair}), and (\ref{H1}) based on the electronic structure 
calculations of a two-dimensional square lattice.  
From the spin-singlet metastable states of the electron pairs, together with the 
single-electron states in the system, we have obtained quantitatively the interaction 
potentials, such as pair-pair interaction and single-electron-electron-pair interaction (both for scattering and transformation), presented in the Hamiltonian starting from the electron-electron Coulomb potential. 

We consider now another possible scenario for the single-electron band. 
In this scenario, the single-electron band is considered as the highest valence band 
of an insulator. The conduction carriers in this band are holes with positive charge 
and dispersion relation $E^h_{{\bf k}}$. For an intrinsic insulator, this band is fully 
occupied by electrons at low temperature and consequently the hole density is zero. 
If we assume that there are $N^{(h)}_F$ holes (from doping initially) in the valence band 
and $N_B$ electron pairs in the electron-pair band (and $N^{(h)}_{F} > 2N_B$),
the total carrier charge number (i.e., the net hole number) in the system is given by $N_h=N^{(h)}_{F}-2N_B$. Transition of two electrons from the valence band to the electron-pair
band creates an electron pair leaving two holes in the valence band.  
The scenario is shown schematically in Fig. 2.
Taking the top of the valence band as the reference for energy,  
the many-particle Hamiltonian in this scenario can be written as,
\begin{equation}\label{Hh}
\tilde{H}^{(v)} = \tilde{H}^{(v)}_0 + \tilde{H}^{(v)}_1,
\end{equation}
with
\begin{eqnarray}\label{Hpairh}
\tilde{H}^{(v)}_0=&& 
\sum_{{\bf k},\sigma} E^h_{{\bf k}} d^{\dag}_{{\bf k},\sigma} d_{{\bf k},\sigma} \nonumber \\
+ && \frac{1}{2A}\sum_{\bf k_1,k_2, q} \sum_{\sigma, \sigma^\prime} v_{q} 
d^{\dag}_{\bf k_1 -q,\sigma} d^{\dag}_{\bf k_2 +q,\sigma^\prime} 
d_{\bf k_2,\sigma^\prime} d_{\bf k_1,\sigma}     \nonumber \\
+ && \sum_{\bf k} \left( E^{\rm pair}_{\bf k}+2E_g\right) b^{\dag}_{\bf k} b_{\bf k} \nonumber \\
+ && \frac{1}{2A}\sum_{\bf k_1,k_2, q} v_{pp}(q) 
b^{\dag}_{\bf k_1 -q} b^{\dag}_{\bf k_2 +q} b_{\bf k_2}b_{\bf k_1}  \nonumber \\
+ && \frac{1}{2A}\sum_{\bf k, k_1, q,\sigma} v^{\rm s}_{hp}(q) 
b^{\dag}_{\bf k_1 -q} d^{\dag}_{\bf k +q,\sigma} b_{\bf k_1} d_{\bf k,\sigma} ,
\end{eqnarray}
and 
\begin{equation}\label{H1h}
\tilde{H}^{(v)}_1= \frac{1}{\sqrt{A}}\sum_{\bf k, q} v^{\rm t}(q)
\left( b^{\dag}_{\bf k } d^{\dag}_{\frac{\bf k}{2}+{\bf q},\uparrow} 
d^{\dag}_{\frac{\bf k}{2}-{\bf q},\downarrow} 
+ b_{\bf k } d_{\frac{\bf k}{2}-{\bf q},\downarrow} 
d_{\frac{\bf k}{2}+{\bf q},\uparrow}  \right), \\
\end{equation}
where the operators $d_{{\bf k},\sigma}^\dag$ and $d_{{\bf k},\sigma}$ are creation and annihilation operators, respectively, for a hole of momentum $\hbar${\bf k} and spin ${\sigma}$. 
They obey the fermion anti-commutation relations
$\{ d_{{\bf k},\sigma}, d^{\dag}_{{\bf k}^{\prime},{\sigma}^{\prime} } \} = 
\delta_{{\bf k},{\bf k}^{\prime}} \delta_{{\sigma},{\sigma}^{\prime}} $,
$\{ d_{{\bf k},\sigma}, d_{{\bf k}^{\prime},{\sigma}^{\prime} } \} = 0$, and
$\{ d^\dag_{{\bf k},\sigma}, d^{\dag}_{{\bf k}^{\prime},{\sigma}^{\prime} } \} = 0$. 
The hole-hole interaction potential $v_q$ is the same as that of the electron-electron interaction. 
The pair-pair interaction potential is also the same as that in Eq.~(\ref{Hpair}), but the hole-electron-pair interaction (scattering) potential is different from the corresponding single-electron-electron-pair potential by a sign, $v^{\rm s}_{hp}(q)= - v^{\rm s}_{ep}(q)$, because the hole has positive charge. The corresponding form factors are also given by Eqs.~(8) and (9).
The second part $\tilde{H}^{(v)}_1$ of the total Hamiltonian is equivalent to $\tilde{H}_1$ given in Eq.(\ref{H1}). It is responsible for 
creating (destructing) an electron pair in the electron-pair band and two holes in the valence band 
at the same time. From the definition of holes, hole creation (annihilation) is actually electron annihilation (creation) in the valence band. So the creation of an electron pair in the electron-pair band and two holes in the valence band is equivalent to the transition of two single electrons from the valence band to the electron-pair band, and the interaction potential $v^t(q)$ keeps the same form giving by Eq.(12).       

The total (net hole) number operator is given by
\begin{eqnarray}
\tilde{N}_h =
\sum_{\bf k,\sigma} d^{\dag}_{{\bf k},\sigma} d_{{\bf k},\sigma}
   - 2\sum_{\bf k} b^{\dag}_{{\bf k}} b_{{\bf k}} .
\end{eqnarray}
It commutates with the total Hamiltonian Eq.~(\ref{Hh}) and is conserved.  

\begin{figure}[htb!]
      {\includegraphics[width=8cm,height=6cm]{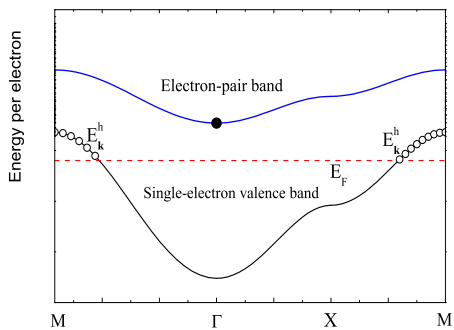}}
       \caption{A schematic plot of the energy band structure with electron pairs and holes.}
       \label{figBGR}
\end{figure}

In this scenario, the stabilization of the electron pairs can be understood as following.
At low density, the attraction between electron pairs and holes are dominant leading 
to the formation of exciton-type states.\cite{AAR} 
The most possible scenario should be an electron pair binds two holes into a biexciton
($X_{2}$).
In two dimensions, the energy of an exciton (for $m_e= m_h$) is approximately $E^{\rm X}_b= - 2{\rm R_y}$. The energy of a biexciton is given by $E^{\rm X_2}_b=2 E^{\rm X}_b + \Delta E^{\rm X_2}_b$, where $\Delta E^{\rm X_2}_b$ is the binding energy
of the biexciton in relation to two free excitons. In a 2D system, this binding energy is about 20\% of the exciton energy $E^{\rm X}_b$.\cite{BX}
Therefore, the energy of an electron pair can be reduced with about 4.4 R$_y$. In other words,
the gap energy $E_g$ can be reduced by 2.2 R$_y$.
At high density, the system tends to be an electron-pair and hole plasma as soon as the electron pairs can stabilized through many-particle interactions. 
Because the electron pairs and holes cannot bind together due to strong interaction or screening of many particles, in this case, the band-gap renormalization due to pair-pair and hole-hole interactions are essential to stabilize the electron pairs. Band-gap renormalization (BGR) has been extensively studied in nonlinear optics of semiconductors, where one
deals with an electron-hole plasma (EHP). In an EHP, the BGR is given by a sum of electron and hole self-energies and is a function of the particle density and temperature.\cite{PV,GT,SDS} It is known that the BGR at finite density can be several times the exciton binding energy in a 2D system. For a high density EHP, the BGR is mainly induced by the Coulomb repulsion among the particles (the Coulomb hole effect). Therefore, we believe that the BGR in our present electron-pair-hole
system should be of similar value, i.e., the renormalization of the electron-pair energy will be enhanced at high density and can be several times the biexciton energy. Consequently, 
the band-gap normalization in the present system can be larger than the single particle gap $E_g$
leading to a negative gap. As soon as the bottom of the electron-pair band touches the Fermi energy $E_F$ of the valence band, a macroscopic electron-pair density is stabilized in the system.  

In summary, we present a Hamiltonian for a 2D electron system with single electrons 
and electron pairs. The electron-pair states are metastable at low electron density
and they are expected to be stabilized through many-particle interactions. 
We have considered two different scenarios for the
single-electron band. In the first case, the single-electron band is considered as the lowest conduction band. In comparison with the FL Hamiltonian, extra terms such as single-electron-electron-pair and pair-pair interactions appear in our Hamiltonian. The interaction potentials 
in the obtained Hamiltonian are determined from the electro-electron Coulomb interaction. 
In the second scenario, the single-electron band is considered as the highest valence band.
We make a connection of the electron pairing and its stabilization to hole density in the system. 
Our present study has shown basic insights into possible electron paring mechanism in a crystal.

\acknowledgments
This work was supported by FAPESP and CNPq (Brazil).

\end{document}